\documentclass[aps,prb,showpacs,twocolumn,showpacs]{revtex4-1}

\usepackage{graphicx}
\usepackage{amsmath}
\usepackage{amssymb}
\usepackage{wasysym}
\usepackage{textcomp}
\usepackage{color}

\definecolor{lila}{rgb}{0.5,0,1}
\newcommand{\bnen}{\begin{equation}}
\newcommand{\eden}{\end{equation}}
\newcommand{\bean}{\begin{eqnarray}}
\newcommand{\eean}{\end{eqnarray}}

\newcommand{\bna}{\begin{array}}
\newcommand{\eda}{\end{array}}

\begin{document}

\title{Entanglement, excitations and correlation effects\\ in narrow zigzag graphene nanoribbons}
\author{I. Hagym\'asi\thanks{hagymasi.imre@wigner.mta.hu} }
\author{\"O. Legeza}
\affiliation{Strongly Correlated Systems "Lend\"ulet" Research Group, Institute for Solid State
Physics and Optics, MTA Wigner Research Centre for Physics, Budapest H-1525 P.O. Box 49, Hungary}

\begin{abstract}
We investigate the low-lying excitation spectrum and ground-state properties of narrow graphene nanoribbons with 
zigzag edge 
configurations. Nanoribbons of comparable widths have been synthesized very recently [P. Ruffieux, \emph{et al.} Nature \textbf{531}, 489 (2016)], and their descriptions
require more sophisticated methods since in this regime conventional methods, like mean-field or density-functional theory with local density approximation, fail to capture the enhanced quantum fluctuations.
Using the unbiased density-matrix renormalization group algorithm we 
calculate the charge gaps with high accuracy for different widths and interaction strengths and compare them
with 
mean-field results. It turns out that the gaps are much smaller in the former case due to the 
proper treatment of quantum fluctuations.
Applying the elements of quantum information theory we also
reveal the entanglement structure inside a ribbon and examine the spectrum of subsystem
density matrices to understand the origin of entanglement. We examine the possibility of
magnetic ordering and the effect of magnetic field.
Our findings are relevant for understanding the gap values in different recent
experiments and the deviations between them.
\end{abstract}
\maketitle
\section{Introduction}
Graphene, the two-dimensional honeycomb lattice of carbon atoms has attracted an enormous interest  
since its first discovery in 2004. \cite{Novoselov666} In spite of this, from the point of view of 
applications in nanoelectronics, bulk graphene is not useful due to the absence of a band gap.
Therefore, finite samples of graphene are likely to be more advantageous in this aspect, since they may
exhibit a gap due to quantum confinement or electronic correlations.
Graphene nanoribbons are especially promising candidates in 
overcoming this obstacle.
 It has been demonstrated, that nanoribbons with a well-defined crystallographic orientation can be produced with
scanning-tunneling-microscope-based litography \cite{Tapaszto:litography,Magda2014} and even sub 4 nm widths can be 
achieved. On the other hand bottom-up techniques now make it possible to synthesize 
either armchair \cite{Kimouche2015} or zigzag \cite{Ruffieux2016} nanoribbons whose widths consist of 
a few zigzag carbon lines only. Ribbons with a zigzag edge are particularly interesting because of 
their peculiar electronic 
and magnetic  properties. \cite{zz_edge_states,louie:prl2006}
While a graphene sheet can be considered as a marginal
Fermi liquid,\cite{graphene:Fermi} and can be treated practically as a non-interacting system,
the situation is completely different for nanoribbons with a zigzag edge. It is known from 
conventional band theory that these ribbons have a flat band
due to their edge states. \cite{zz_edge_states} This large density of states at the Fermi energy is 
very sensitive to magnetic
ordering, even if only a weak electron-electron interaction is present based on the the Slater theory of 
antiferromagnetism.   Note that such a 
drastic effect does not occur in armchair ribbons due to the absent edge states, therefore
we focus on zigzag ribbons in the following.
As a result of the interaction, a gap opens in zigzag ribbons, which implies magnetically ordered edge states as it has been demonstrated  in an 
indirect way recently with the contribution of 
one of us. \cite{Magda2014}
\par The above facts motivated the exploration of the interaction 
effects with the use of several methods, like density-functional theory (DFT), 
\cite{louie:prl2006,Louie:GW,PhysRevLett.101.096402,Katsnelson:DFT,Chen:DFT} 
mean-field approximation,
\cite{Rossier:HF,Yazyev:prl2008,MacDonald:prb2009,Pi:HF2015,Jung:HF,Jung:HFprl,Hirashima:HF,
Lewenkopf:HF}
quantum Monte 
Carlo (QMC) \cite{Feldner:prb2010,Feldner:prl2011,Schmidt:QMC} and density-matrix  renormalization 
group 
algorithm (DMRG). \cite{Hikihara:dmrg2003,Mizukami:dmrg,Ramasesha:gnr}
The DFT and DMRG in Ref.~[\onlinecite{Mizukami:dmrg}] are used in ab-initio calculations,  while the 
other methods are applied in solving 
the $\pi$-band model of graphene, that is, the Hubbard model on a honeycomb ribbon:
\begin{equation}
\label{eq:Hamiltonian}
\mathcal{H}=\sum_{ij}t_{ij}\hat{c}^{\dagger}_i\hat{c}^{\phantom\dagger}_j+U\sum_i\hat{n}_{i\uparrow}
 \hat{n}_{i\downarrow},
\end{equation}
where $t_{ij}$ is the hopping amplitude between sites $i$ and $j$, and $U$ is the strength of the 
local Coulomb interaction. In what follows we consider only nearest-neighbor hopping
terms, with $t=2.7$ eV.
The most extensive studies
were performed with DFT and in mean-field approximation because broad ribbons can easily
be accessed, while alternative approaches have been suggested recently to treat
large 
system sizes. \cite{Schmidt:prb} Density functional and mean-field theories reduce the 
original interacting system to an effective single-particle problem,  therefore they neglect 
correlation effects and quantum fluctuations which are known to be significant in quasi one-dimensional
systems. For this reason a more accurate description is necessary,  which QMC and DMRG
are able to address. Due to the exponential growth of
the Hilbert space, these methods can treat only much smaller sizes.  The QMC studies mainly focused 
on
the correlation between edge atoms and dynamical properties, \cite{Feldner:prl2011} and the  
previous DMRG model calculations were 
restricted to very small systems. \cite{Hikihara:dmrg2003,Ramasesha:gnr}
In the widespread use of mean-field theory, it is important to investigate its reliability by 
examining  the role of enhanced quantum fluctuations, which this approximation neglects. 
Although a benchmark of mean-field theory was performed for quantum dot-like 
structures, \cite{Feldner:prb2010}
less is known about the case of zigzag nanoribbons. It has been shown previously that  QMC results 
agree  well quantitatively with mean-field results for wide enough ribbons in the weakly interacting 
limit. \cite{Feldner:prl2011} In light of the new fact that zigzag ribbons with already five 
zigzag carbon lines width can be created, \cite{Ruffieux2016} it is necessary to analyze such 
narrow ribbons where 
dimensionality effects are  expected to be more crucial.
 Here we intend to
fill a major gap with the DMRG method by investigating the low-lying energy spectrum of zigzag ribbons in a 
controlled, accurate manner. We calculate the charge gaps for various interaction
strengths and compare them with the mean-field results. Moreover, they are also experimentally 
relevant  quantities, since they can be accessed by
scanning tunneling microscopy measurements. \cite{Magda2014}
Furthermore, we
address the ground-state properties by considering not only the correlations between the edge atoms, 
 but between every pair of sites, which
provides us a deeper insight into the many-body aspect of
the graphene nanoribbons. We also investigate the magnetic properties of the ground state and the 
effect of external magnetic field.
\par The paper is organized as follows. Section II.~ contains the numerical details of the DMRG 
method.
In Section III.~A we present our results for the charge gaps for various ribbon widths and 
interaction strengths 
and compare them with the mean-field results. Section III.~B demonstrates the application of 
quantum 
information theory in determining the entanglement patterns and correlation functions of 
nanoribbons. In Section III.~C the magnetic properties are addressed.
In Section III.~D we discuss how our results for the charge gap are related to recent experiments 
and 
DFT calculations. Finally, in 
Section IV. our conclusions 
are presented.

\section{Methods}
\begin{figure}[!htb]
\includegraphics[width=0.7\columnwidth]{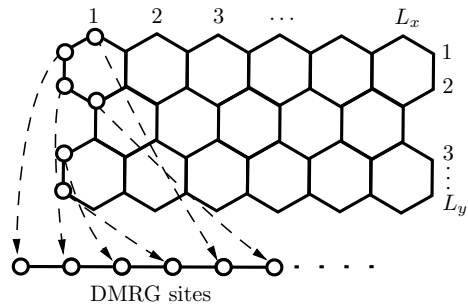}
\caption{The applied notations for the length and width of a zigzag ribbon, and the mapping used in
the DMRG calculation.}
\label{fig:mapping}
\end{figure}

 We apply the DMRG algorithm in real 
space \cite{White:DMRG1,White:DMRG2,schollwock2005,manmana2005,hallberg2006,legeza:review}
and use the dynamic 
block-state selection approach
(DBSS), \cite{DBSS:cikk1,DBSS:cikk2} which enables us the accurate control of the truncation error. 
In our case,  the a priori value 
of the quantum information loss was set 
to $\chi=10^{-4}$ and the truncation errors were in the order of $10^{-6}$.  This threshold value required 
block states about 15000-20000 so that, our results are 
far more accurate than those from previous investigations,
\cite{Hikihara:dmrg2003,Mizukami:dmrg,Ramasesha:gnr} in terms of the truncation procedure.
Such a large number of block states is necessary to obtain accurate gap values and correlation functions. 
We considered ribbons with a maximum number of sites $L=84$ and open boundary condition
is applied at every edge. 
A honeycomb 
ribbon is mapped to a one-dimensional chain, with long-range hopping elements. The ordering of the 
sites and the geometry of the ribbon is given in Fig.~\ref{fig:mapping}.
\section{Results}
\subsection{Charge gap}
\begin{figure}[!htb]
\includegraphics[width=0.8\columnwidth]{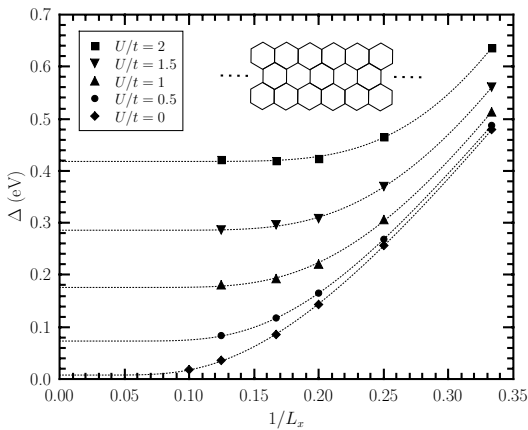}
\caption{Finite-size scaling of charge gaps for different values of $U$ and $L_y=4$ as
indicated
by the inset figure. The dotted lines denote the exponential fit to the data described in the main text.}
\label{fig:gap_scaling}
\end{figure}

\begin{figure}[!htb]
\includegraphics[width=0.8\columnwidth]{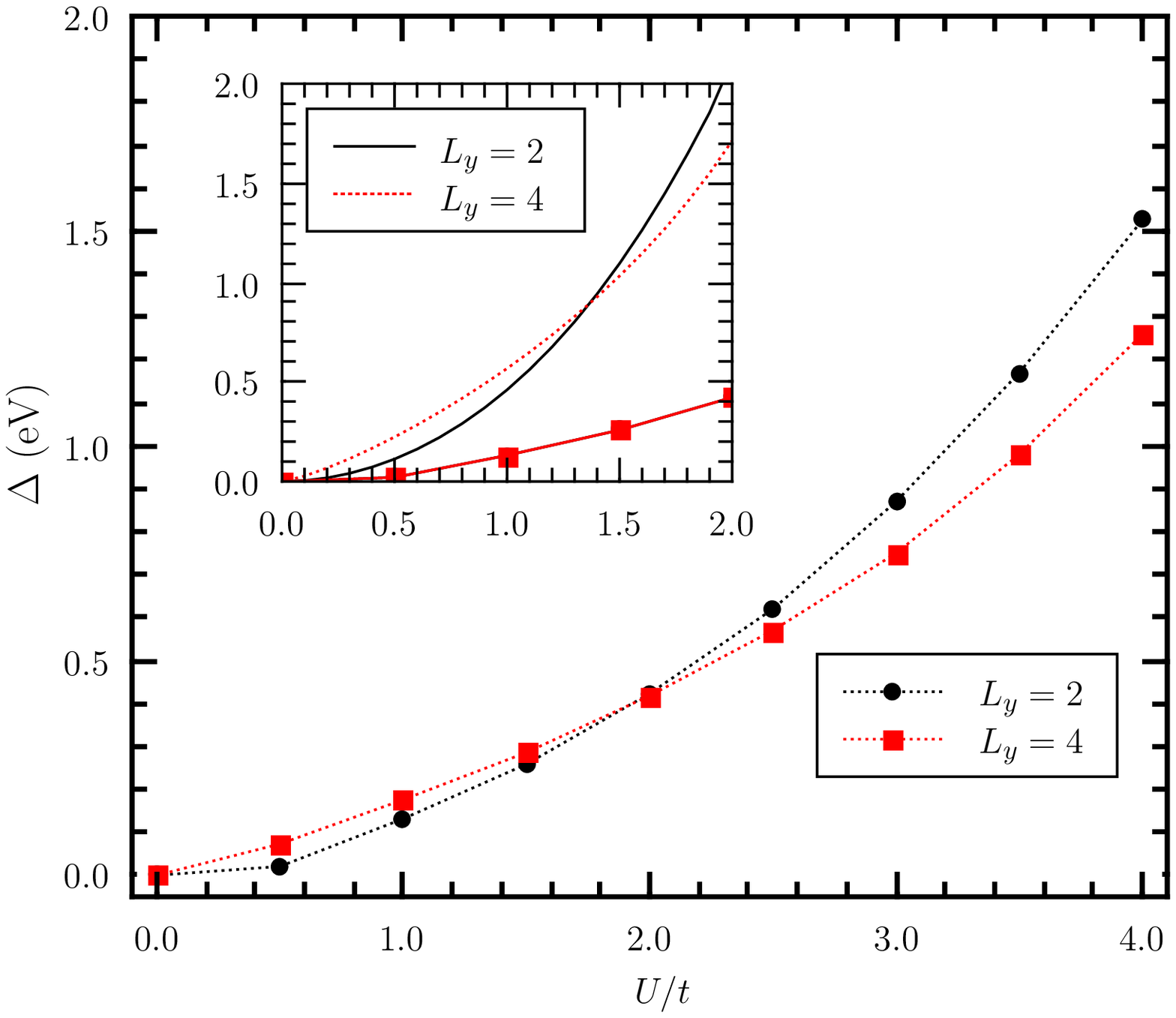}
\caption{(color online) DMRG results for the charge gaps as a function of $U$ for two different widths
($L_y=2$ and 4) as indicated
in the legend. The dotted lines are guides to the eye. The inset shows the bandgaps obtained with 
the mean-field approximation of the Hubbard-model (solid and dotted lines) together with the DMRG 
data.}
\label{fig:gap_U}
\end{figure}

In what follows we restrict ourselves to half-filled case. We know that in this case Lieb's 
theorem \cite{Lieb:prb} forbids the appearance
of spontaneous spin polarization, however, the low-lying spectrum can be obtained. In fact, with scanning tunneling
microscopy one can measure the band gap, which in our case corresponds to the single-particle excitation,
usually referred as the charge gap:  
\begin{align}
\Delta(L)&=\frac{1}{2}\left[E_0\left(\frac{L}{2}+1,\frac{L}{2}\right)+E_0\left(\frac{L}{2}-1,\frac{L
}{2}\right)\right.\nonumber\\
&\left.-2E_0\left(\frac{L}{2},\frac{L}{2}\right)\right],
\end{align}
where $E_0(N_{\uparrow},N_{\downarrow})$ is the ground state in sector with $N_{\uparrow}$ up-spin
and $N_{\downarrow}$ down-spin electrons and $L=2L_xL_y+L_y$ is the total number of sites. 
We calculated the charge gaps for different values of $U$ and two different widths,
$L_y=2$ and $4$. The maximum length we could achieve was $L_x=10$ for $L_y=4$. The
Hubbard $U$ was varied within the range $U/t=0$ and 4, since around $U_c/t\sim3.9$ a 
Mott-transition occurs in  the two-dimensional honeycomb lattice, 
\cite{Mott1:prx,Mott2:prb,Mott3:prx}
and larger values are not physical in case of graphene. 
A careful extrapolation of the gaps to the thermodynamic limit was performed in each case as it is demonstrated in
Fig.~\ref{fig:gap_scaling} for $L_y=4$. 
The data for $L_y=4$ were fitted using an exponential function:
\begin{align}
\Delta(L_x)=\Delta(\infty)+A\exp\left(-B/L_x\right),
\end{align}
with $\Delta(\infty)$, $A$ and $B$ being free parameters. For $L_y=2$ we obtained the same results 
as in Ref.~[\onlinecite{Hikihara:dmrg2003}] (see Fig.~\ref{fig:gap_U}), where the finite-size scaling of the gaps can be 
fitted 
with a quadratic function, which may be due to the stronger one-dimensional effects. In this case a 
much smaller number of block states, $\sim 500$, was sufficient to keep the truncation errors in 
the order of $10^{-6}$. However, for $L_y=4$ 15000-20000 block states are necessary to obtain 
energy values within the same error margin.
Since the ground-state energies are in the order of $\sim10^2$ and the magnitude of the truncation
errors is $\sim10^{-6}$, we estimate that the error of the gaps is around $\sim 10^{-4}$, which is 
much smaller than
the size of the symbols in Fig.~\ref{fig:gap_scaling}.
The extrapolated gap values as a function of $U$ are shown in Fig.~\ref{fig:gap_U} for two different widths. 
Our results suggest that the charge gap opens at $U/t=0$, which might be the residue of the 
Slater transition occurring in the mean-field treatment. This is also in agreement with the previous 
 prediction \cite{Hikihara:dmrg2003} based on the analysis of narrower ribbons.  It is worth 
comparing  these results to those obtained from mean-field theory whose
details 
have been described in several papers. \cite{MacDonald:prb2009,Feldner:prb2010} We performed 
mean-field calculations for the present ribbons (see Appendix for the details), and the obtained 
mean-field gaps 
are shown in the inset of Fig.~\ref{fig:gap_U}. They exhibit
qualitatively very similar behavior, however, they are remarkably larger than the DMRG gap values. This can be
attributed to the neglect of quantum fluctuations, which are enhanced at such narrow widths. 
It is worth mentioning that for small two-dimensional structures, the mean-field theory provided quite accurate
results in comparison with QMC. \cite{Feldner:prb2010} This may follow from the fact that the 
quantum fluctuations in a two-dimensional  systems are not as strong as in one dimension.
\subsection{Quantum information analysis, correlation functions}
 As a next step we consider the ground-state properties by investigating various correlation 
functions.
Our aim is to investigate the correlations between two given subsystems,  namely, between two 
sites. The knowledge of this
quantity provides information about the whole system. This can be obtained from
the mutual information: \cite{wolf2008,furukawa2009,legeza:entanglement}
\begin{gather}
   I_{ij}=s_i+s_j-s_{ij},
\end{gather}
which measures all types of correlations (both of classical and quantum origin) between sites $i$ 
and $j$, we will refer to this quantity 
as the strength of entanglement between the system components. 
Here $s_i$ and $s_{ij}$ are the one- and two-site von Neumann 
entropies, \cite{legeza2003b,vidallatorre03,calabrese04,legeza2006,rissler2006,luigi2008}
respectively, that can be calculated from the corresponding reduced density matrices:
\begin{align}
 s_i&=-{\rm Tr} \rho_i\ln\rho_i,\\
 s_{ij}&=-{\rm Tr} \rho_{ij}\ln\rho_{ij},
\end{align}
 where $\rho_i$ ($\rho_{ij}$) is the reduced density matrix of site $i$ (sites $i$ and $j$),
which is derived from the density matrix of the total system by tracing out the configurations of
all other sites. 
In this part, we calculate the mutual information for different 
values of the Hubbard interaction to reveal how the interaction
modifies the original ground state.
\begin{figure*}[!ht]
\includegraphics[width=1.6\columnwidth]{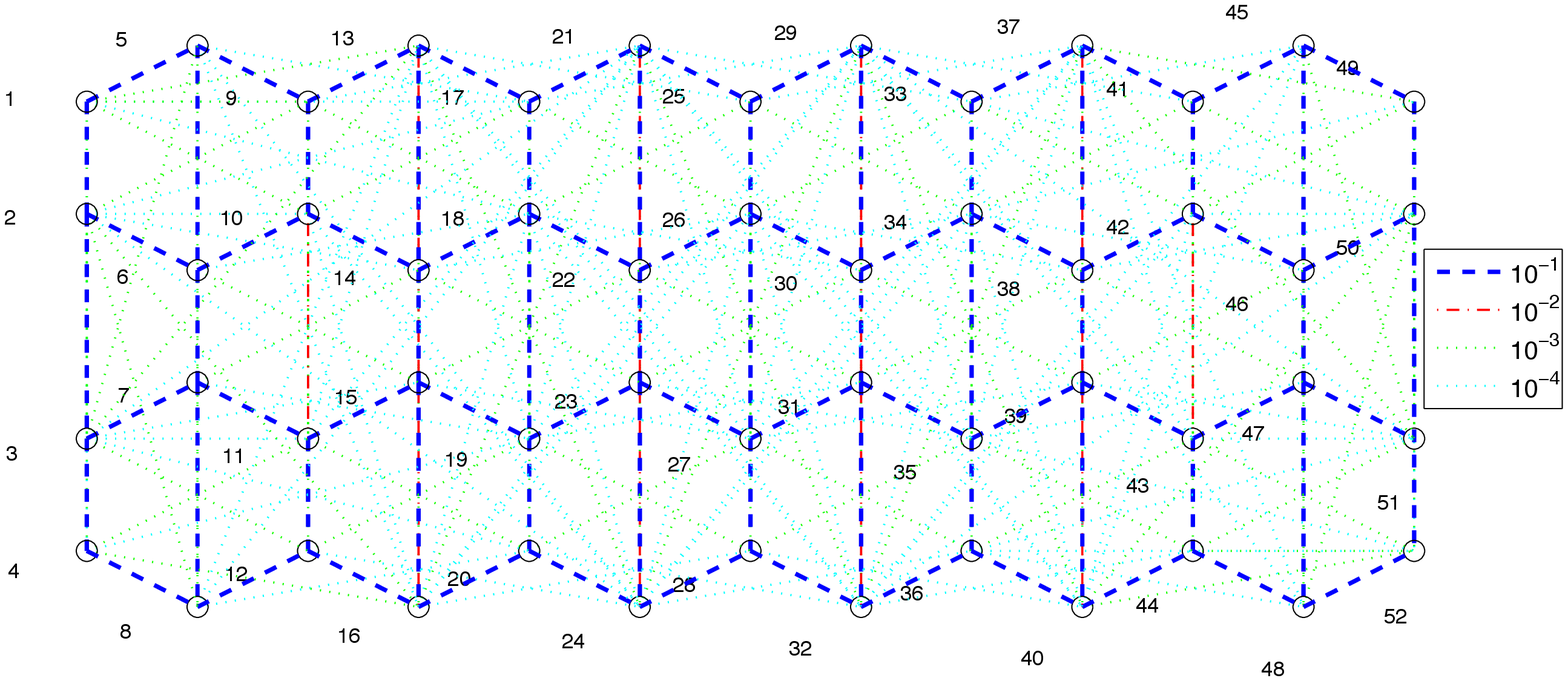}
\caption{(color online) Entanglement patterns in a zigzag ribbon for $L_x=6$, $L_y=4$ and $U=0$. The
various types of lines  
correspond to different magnitudes as indicated in the sidebar. The numbers indicate the positions 
of sites along the one-dimensional DMRG topology. The blue dashed lines connect only 
nearest-neighbor sites (for example: sites $i=1,2$ or $i=1,5$), or opposite sites within a
hexagon 
(for example: sites $i=9,10$ or $i=22,23$). The red dash-dot lines connect opposite zigzag
sites, for 
example: $i=13,16$.}
\label{fig:mutual_inf_U0_zz}
\end{figure*}
\begin{figure*}[!ht]
\includegraphics[width=1.6\columnwidth]{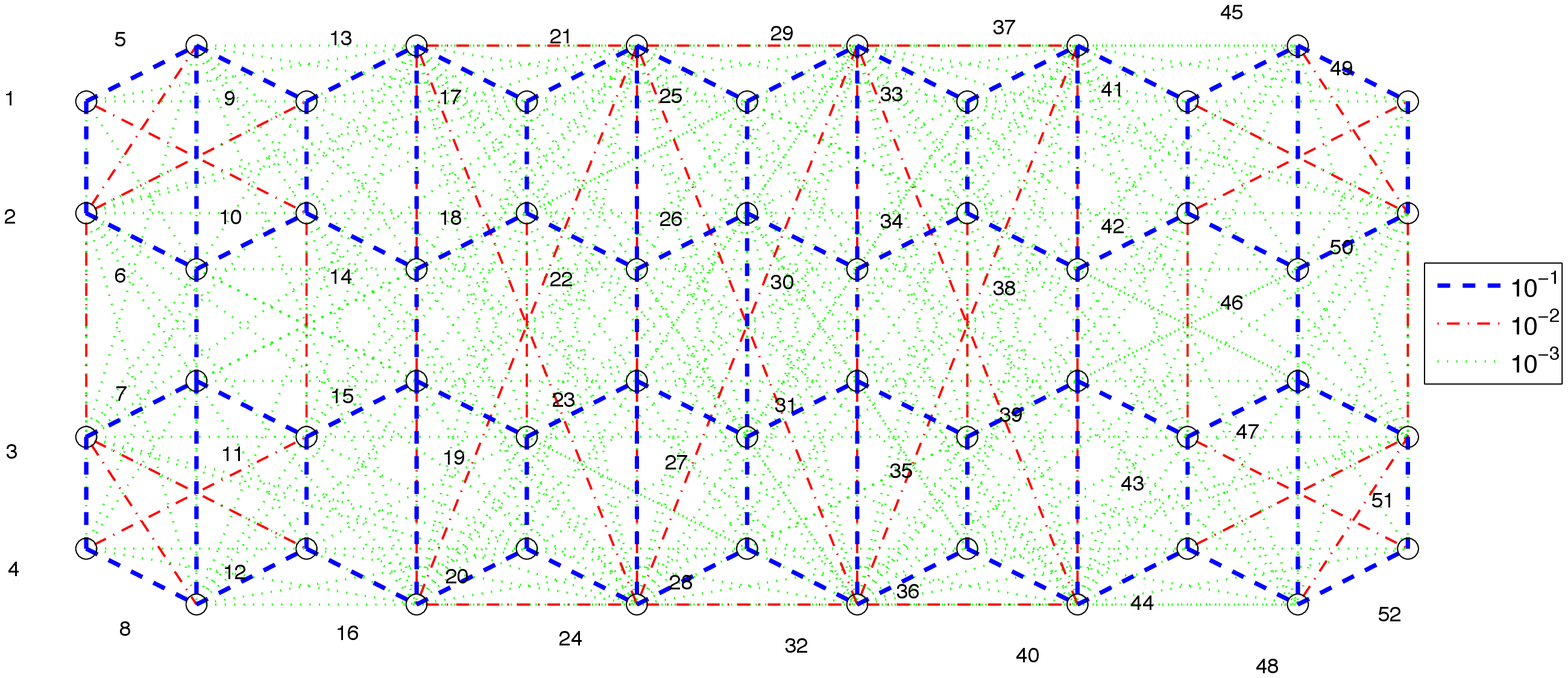}
\caption{(color online) Similar to Fig.~\ref{fig:mutual_inf_U0_zz} but for $U/t=2$, furthermore the 
red dash-dot lines at the zigzag edges connect neighboring zigzag sites, for example:
$i=13,21$ or 
$i=24,32$. }
\label{fig:mutual_inf_U2_zz}
\end{figure*}
We use the sum of one-site entropies,\cite{legeza2004} $I_{\rm TOT}$,  and the entanglement 
distance,\cite{rissler2006} $I^{\rm (MPS/real),\eta}_{\rm dist}$ :
\begin{align}
I_{\rm TOT}&=\sum_is_i,\\
I^{\rm (MPS/real),\eta}_{\rm dist}&=\sum_{ij}I_{ij}\left(d_{ij}^{\rm (MPS/real)}\right)^{\eta},
\end{align}
where $d_{ij}^{\rm(MPS)}=|i-j|$ is for the one-dimensional topology of the DMRG and $d_{ij}^{\rm(real)}$ is the distance
in physical lattice space. $I_{\rm TOT}$ and $I^{\rm (MPS/real),\eta}_{\rm dist}$ quantify the total quantum information encoded in the wave function
and the localization of entanglement in the system, respectively.
Firstly, we consider the noninteracting case. The entanglement  patterns 
obtained from the mutual information are shown in Fig.~\ref{fig:mutual_inf_U0_zz} for a system with $L_x=6$
 and $L_y=4$.
It is clearly observed, that mainly short-range correlations are present, and certain opposite  
sites in a hexagon are entangled, but there is no strong entanglement between the two edges (note that red and green lines correspond to one and two orders of magnitude smaller values, respectively). 
In this case a maximum number
of block states $\sim8000$ was sufficient to determine the ground state wave function within our error margin.
As a next step, we investigate what happens when the electrons are interacting.  To emphasize the 
interaction effects, we set $U/t=2$. The results are shown in Fig.~\ref{fig:mutual_inf_U2_zz}.
It is remarkable, that besides the strong nearest-neighbor entanglement, moderately strong 
entanglement appears between the two edges, and between electrons on the same edge. It is 
interesting to mention that here a maximum number of $\sim15000$ block states was necessary to 
obtain the ground state, which is almost twice as large as in the noninteracting case using the 
same threshold value in the truncation procedure. 
In agreement with this $I^{\rm (MPS),2}_{\rm dist}(U=0)=232.07$ increases  to  $I^{\rm (MPS),2}_{\rm 
dist}(U=2t)=371.12$ and similarly $I^{\rm (real),2}_{\rm dist}(U=0)=38.67$ increases to  $I^{\rm 
(real),2}_{\rm dist}(U=2t)=51.27$.
Therefore, for $U/t=2$ 
longer range entanglement bonds appear, as indicated by $I^{\rm (real),2}_{\rm dist}$, in  contrast 
to the $U=0$ case, whose presence naturally 
requires much larger bond dimensions since these entanglement bonds are cut when subsystem 
entropies are calculated. The sum of one-site entropies decreases  with $U$ ($I_{\rm 
TOT}(U=0)=72.08$, $I_{\rm TOT}(U=2t)=70.45$), but in a much lower rate than
in the one-dimensional case. We also mention that the ground states in both cases are spin 
singlets, and finite spin polarization does not appear at the edges, since the ground state respects 
the rotational symmetry of the original Hamiltonian, in agreement with the previous DMRG and QMC 
results. This is in sharp contrast with the mean-field results, where a broken-symmetry ground state is 
realized and the ferromagnetically polarized edges are coupled to each other antiferromagnetically.
\par The analysis so far has given us an overall picture about which sites are strongly entangled, but to 
obtain additional information about the nature of the entanglement it is worth investigating the 
eigensystem of the two-site density matrices. Firstly, we consider two neighboring zigzag sites, 
(24 and 32 in Fig.~\ref{fig:mutual_inf_U2_zz}) and solve the eigenvalue problem of the 
corresponding two-site reduced density matrix, $\rho_{24,32}$. In its eigenvalue spectrum, the most 
significant eigenvalue ($\omega=0.128$) is threefold degenerate, and the corresponding eigenvectors 
are: \begin{equation}
\label{eq:wavefunction:V0p3_onsite}
\begin{split}
 \phi^{(1)}_{24,32}=& \ |\uparrow\rangle_{24}|\uparrow\rangle_{32},\\
 \phi^{(2)}_{24,32}=& \
\frac{1}{\sqrt{2}}(|\uparrow\rangle_{24}|\downarrow\rangle_{32}+|
\downarrow\rangle_{24}|\uparrow\rangle_{32}\rangle),\\
 \phi^{(3)}_{24,32}=& \ |\downarrow\rangle_{24}|\downarrow\rangle_{32}.
\end{split}
\end{equation}
It means that in this mixed state the largest weight belongs to the triplet components, which 
results in a ferromagnetic correlation between the two neighboring zigzag sites. Similarly, we 
investigate the reduced density matrix of two zigzag sites sitting on opposite edges, e.g. 29 and 
32. Performing the same analysis, we find that the eigenvector corresponding to the largest 
eigenvalue ($\omega=0.19$) is:
\begin{equation}
\label{eq:wavefunction:V0p8_nn}
\begin{split}
 &\phi_{29,32}=\\
  &0.7067(|\uparrow\rangle_{29}|\downarrow\rangle_{32}
 -|
\downarrow\rangle_{29}|\uparrow\rangle_{32})\\
& + 0.0236(|\uparrow\downarrow\rangle_{29}|0\rangle_{32}
+|0\rangle_{29}
|\uparrow\downarrow\rangle_{32}),
\end{split}
\end{equation}
which describes that the two electrons on the opposite edges form a singlet.
This can be considered as a direct evidence for the antiferromagnetic coupling between the two 
edges mediated by the conduction electrons in the ribbon. 
\par
\begin{figure}[!t]
\includegraphics[width=0.8\columnwidth]{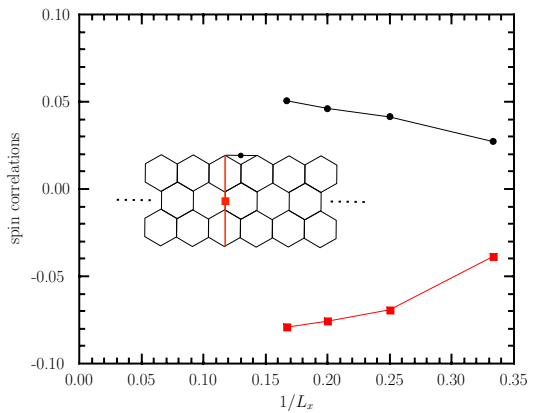}
\caption{(color online) Finite-size scaling of spin correlations between the opposite edges (red 
squares)
and between neighboring edge atoms  for $U/t=2$ (black circles), calculated in the middle
of the ribbon as they are indicated in the inset figure.}
\label{fig:spin_corr}
\end{figure}

 The above statements have been obtained for a finite length, thus it is important to 
investigate their
size-dependence, which is shown in Fig.~\ref{fig:spin_corr}.
Here one can see the spin correlations ($\langle\boldsymbol{S}_i\boldsymbol{S}_j\rangle$) between 
the opposite edges and between neighboring edge atoms as a function of inverse ribbon length taken 
at the middle
of the ribbon. In the case of 
odd $L_x$ values we performed an average over the two correlation values in the middle of the ribbon 
to reduce the oscillation due to the finite-size effects.
It is easily seen that the absolute value of both quantities increases with the ribbon length, 
confirming that
the above results remain valid even in the thermodynamic limit.

\subsection{Magnetic properties}
\begin{figure*}[!htb]
\includegraphics[width=1.6\columnwidth]{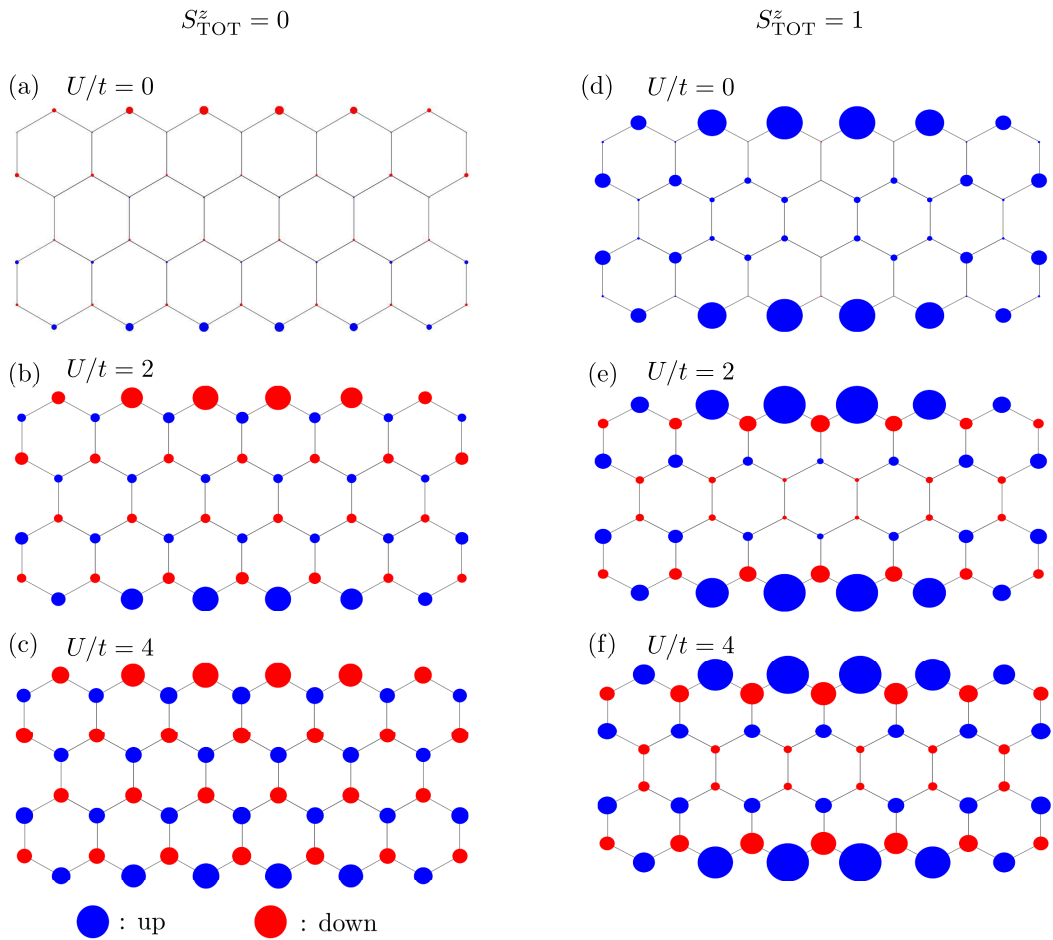}
\caption{(color online) Panels (a)-(c) show the local magnetic moments ($S_i^z$) of the ground 
state in the presence of a pinning magnetic field at the bottom zigzag sites, $h/t=0.01$, for 
various values of $U$. Panels (d)-(f) show the distribution of local magnetic moments in the 
$S_{\rm TOT}^z=1$, $S=1$ sector for different values of $U$. The magnitude of up and down moments 
are proportional to the area of the circles.   }
\label{fig:ribbon_magnetic}
\end{figure*}

Previously we revealed the behavior of correlation functions inside the ribbon. Naturally, we did 
not find long-range magnetic order in finite-systems due to the SU(2) symmetry of the Hamiltonian. 
However, this may not be true in the thermodynamic limit, where symmetry-breaking ground
state can 
occur.\cite{Schmidt:QMC} We investigate this aspect, by adding an artificial
pinning magnetic 
field along the $z$-direction at the bottom sites of the ribbon to the Hamiltonian
(\ref{eq:Hamiltonian}):
\begin{equation}
 \mathcal{H}_{\rm pin}=-h\sideset{}{'}\sum_iS_i^z,
\end{equation}
where the prime denotes that the summation is over only the bottom sites of the ribbon (for example 
sites 8,16,24,$\dots$ in Fig.~\ref{fig:mutual_inf_U0_zz}). We apply a tiny magnetic field, 
$h=0.01t$, to explore possible magnetic order by investigating the response of the system for 
various values of the Hubbard interaction. The results are shown in Figs.~\ref{fig:ribbon_magnetic} 
(a)-(c). 
One can clearly see in Fig.~\ref{fig:ribbon_magnetic} (a) that for $U/t=0$ the pinning magnetic 
field hardly triggers any magnetic moment, as it is expected for a paramagnetic ground state. The 
situation is quite different as $U$ is switched on. For $U/t=2$ remarkable spin polarization 
appears at the edges, $S_i^z\sim 0.06$, while much smaller magnetic moments appear inside the 
ribbon. This corresponds to the regime where the edge magnetism is expected to occur. Increasing 
$U$ near the critical $U_c$, significant spin polarization appears also inside the ribbon, whose 
magnitude is comparable to the magnetic moments at the edges.  
This reflects the tendency that the 
honeycomb lattice becomes antiferromagnetically ordered above the critical value. However,
here a crossover takes place since the charge gap opens up for any $U/t>0$ in the zigzag
ribbon unlike in the fully two-dimensional 
honeycomb lattice, where the Mott transition occurs at a finite $U_c$.
\par As a next step we investigate what happens if a uniform external magnetic field is applied. To 
address this question we consider the first spin excited state of the Hamiltonian
(\ref{eq:Hamiltonian}), with $S_{\rm TOT}^z=1$, $S=1$ 
quantum numbers. The results are shown in Figs.~\ref{fig:ribbon_magnetic} (d)-(f).
Qualitatively similar behavior is observed in the bulk as in the previous case, 
however, the polarization of the edges is more robust in all cases, even for $U/t=4$. This 
substantiates the findings of Ref.~[\onlinecite{Hikihara:dmrg2003}] that the zigzag sites at the 
edges can be polarized in the easiest way.

\subsection{Discussion}
 We discuss our results in the light of recent experiments \cite{Ruffieux2016}
where ribbons with a comparable width were investigated.
To account for the reported gaps of \mbox{$\Delta_{\rm exp}\sim 1.5\, {\rm eV}$}, the 
Hubbard-$U$ should be tuned very close to the Mott-insulating
regime, see Fig.~\ref{fig:gap_U}. This may not be a surprise for us since in 
free-standing graphene the Hubbard-$U$ was estimated $U/t\sim
3.4$,\cite{graphene:abinitio} which is quite close to $U_c$. 
Larger values of $U$ are not reasonable in our case, since
the bulk graphene is not an insulator.
Note that the gaps may be further increased by the inclusion of longer-range interactions,
since the screening may not
be as effective as in ordinary metals.
This hypothesis is corroborated by DFT+$GW$ 
calculations, \cite{Louie:GW,dft:substrate} where a significant 
increase of the gaps has been observed compared with what has been obtained in local-density 
approximation (LDA). \cite{louie:prl2006} It is also in agreement with the fact that the 
scanning-tunneling-microscope measurements were carried out in such an environment
where the sample was embedded on an insulator. \cite{Ruffieux2016} Here one can expect that
the on-site Coulomb interaction is close to the value of the free-standing graphene and
long-range interaction might be important. Thus, as we have seen $U/t\sim 3.4-3.9$
results in gaps lying very close the experimental values.
On the other hand, when the ribbon is placed on the top of a metal host, \cite{Magda2014}
the experimental gap values, $\Delta_{\rm exp}\sim 0.3$ eV, are smaller by almost an order
of
magnitude. This can be explained,
if we recall that in this case the host metal induces extra charge carriers into the
ribbon -- indicated by the presence of
a finite density of states at the Fermi energy --, therefore the screening of the Coulomb
interaction is stronger. Hence,
the $\pi$-band model
containing a much weaker Hubbard interaction term,
$U/t\sim1.5$, than in the free-standing case, provides quantitatively accurate gap values
in
this case. Another possible importance of our results is the fact that the gap values cannot be increased
above $\Delta\sim0.3$-$0.4$ eV
by creating narrower and narrower ribbons -- in contrast to prediction of the naive mean-field theory --, as long as the
ribbon can be described by the conventional $\pi$-band model, like in the experimental setup of 
Ref.~[\onlinecite{Magda2014}]. Furthermore,  we emphasize that the widely used mean-field
approach cannot be applied even for wider ribbons for $U/t\gtrsim2.23$, since this theory
predicts a Mott-insulating state in the honeycomb lattice above this
value,\cite{Tosatti:epl} which turned out to be inaccurate by more sophisticated
calculations.\cite{Mott1:prx,Mott2:prb,Mott3:prx}
\section{Conclusions}
In this paper we examined the charge gaps and ground-state properties of narrow zigzag graphene 
nanoribbons by applying the unbiased DMRG method with high accuracy to the $\pi$-band model containing only local interaction terms. 
It turned out that the mean-field theory grossly overestimates the gap values in the case
of such narrow
widths. This discrepancy can be ascribed to the fact that the enhanced quantum fluctuations suppress 
their values. 
Our analysis also revealed how the deviations between recent experiments can be understood
in terms of the $\pi$-band model with a
local Coulomb interaction. It was argued that depending on the effectiveness of screening,
tuning the value of the on-site Hubbard term, the model can account for quantitatively accurate gap
values. 
We performed a quantum information analysis and
determined the spectrum of subsystem density matrices and the entanglement patterns in the ribbons, which gave us a spectacular description of the 
many-body aspect of the ground states. We pointed out how the entanglement evolves as the 
interaction is switched on. 
The understanding of the entanglement structure in nanoribbons is important from the point of view of their future
applications in quantum information processing or quantum computation.
Finally, we investigated the magnetic properties of nanoribbons and explored possible
magnetic orders for various values of the Hubbard interaction.
\acknowledgements{ We acknowledge helpful discussions with F. Gebhard, L. Tapaszt\'o and P. 
Vancs\'o.
This work was supported in part by the
Hungarian Research Fund (OTKA) through Grant Nos.~K120569 and NN110360.}

\appendix*
\section{Details of the mean-field calculation}
We performed mean-field calculations to benchmark its gap values against the DMRG
results. Using the standard procedure for the decoupling of the Hubbard term we arrive at
\begin{align}
 \mathcal{H}^{'}=&-t\sum_{\langle ij \rangle\sigma}\left(
\hat{c}_{i\sigma}^{\dagger}\hat{c}_{j\sigma}^{\phantom\dagger}+\textrm{H.c.}
\right)\nonumber\\
&+U\sum_i\left(\hat{n}_{i\uparrow}\langle \hat{n}_{i\downarrow}
\rangle+\hat{n}_{i\downarrow}\langle \hat{n}_{i\uparrow} \rangle -\langle
\hat{n}_{i\downarrow}
\rangle \langle \hat{n}_{i\uparrow} \rangle \right),
\end{align}
where we have made use of the fact that only nearest-neighbor hoppings are allowed and
the summation in the first term is carried out for nearest-neighbor sites. Since we deal
with a single-particle problem, the Hamiltonian can be diagonalized by a Bogoliubov
transformation in $k$-space:
\begin{align} 
\mathcal{H}^{'}=\sum_{k\sigma n}\varepsilon_{nk\sigma}\hat{C}^{\dagger}
_ { n k
\sigma }\hat{C}^{\phantom\dagger}_{nk
\sigma
}
-U\sum_i\langle\hat{n}_{i\uparrow}\rangle\langle\hat{n}_{i\downarrow}\rangle,
\end{align}
where $\hat{C}^{\dagger}_{nk\sigma}$
($\hat{C}^{\phantom\dagger}_{nk\sigma}$) are the transformed operators that
destroy (create) a particle with wavenumber $k$ with spin $\sigma$ in band $n$. The
energy bands are given by $\varepsilon_{nk\sigma}$ which depend on the yet unknown
electron densities. The densities and energy bands are calculated selfconsistently. For a
given ribbon width, $L_y$, we obtain $2L_y$ bands according to the number of sites in the
unit cell of the ribbon. Since we deal with the half-filled case, the first $L_y$ bands
are completely filled, thus, the energy gap is determined as the bandgap between bands
$n=L_y$ and $n=L_y+1$. This is shown in the inset of Fig.~\ref{fig:gap_U}.
\bibliography{paper_graphene} 

\begin{thebibliography}{52}%
\makeatletter
\providecommand \@ifxundefined [1]{%
 \@ifx{#1\undefined}
}%
\providecommand \@ifnum [1]{%
 \ifnum #1\expandafter \@firstoftwo
 \else \expandafter \@secondoftwo
 \fi
}%
\providecommand \@ifx [1]{%
 \ifx #1\expandafter \@firstoftwo
 \else \expandafter \@secondoftwo
 \fi
}%
\providecommand \natexlab [1]{#1}%
\providecommand \enquote  [1]{``#1''}%
\providecommand \bibnamefont  [1]{#1}%
\providecommand \bibfnamefont [1]{#1}%
\providecommand \citenamefont [1]{#1}%
\providecommand \href@noop [0]{\@secondoftwo}%
\providecommand \href [0]{\begingroup \@sanitize@url \@href}%
\providecommand \@href[1]{\@@startlink{#1}\@@href}%
\providecommand \@@href[1]{\endgroup#1\@@endlink}%
\providecommand \@sanitize@url [0]{\catcode `\\12\catcode `\$12\catcode
  `\&12\catcode `\#12\catcode `\^12\catcode `\_12\catcode `\%12\relax}%
\providecommand \@@startlink[1]{}%
\providecommand \@@endlink[0]{}%
\providecommand \url  [0]{\begingroup\@sanitize@url \@url }%
\providecommand \@url [1]{\endgroup\@href {#1}{\urlprefix }}%
\providecommand \urlprefix  [0]{URL }%
\providecommand \Eprint [0]{\href }%
\providecommand \doibase [0]{http://dx.doi.org/}%
\providecommand \selectlanguage [0]{\@gobble}%
\providecommand \bibinfo  [0]{\@secondoftwo}%
\providecommand \bibfield  [0]{\@secondoftwo}%
\providecommand \translation [1]{[#1]}%
\providecommand \BibitemOpen [0]{}%
\providecommand \bibitemStop [0]{}%
\providecommand \bibitemNoStop [0]{.\EOS\space}%
\providecommand \EOS [0]{\spacefactor3000\relax}%
\providecommand \BibitemShut  [1]{\csname bibitem#1\endcsname}%
\let\auto@bib@innerbib\@empty
%</preamble>
\bibitem [{\citenamefont {Novoselov}\ \emph {et~al.}(2004)\citenamefont
  {Novoselov}, \citenamefont {Geim}, \citenamefont {Morozov}, \citenamefont
  {Jiang}, \citenamefont {Zhang}, \citenamefont {Dubonos}, \citenamefont
  {Grigorieva},\ and\ \citenamefont {Firsov}}]{Novoselov666}%
  \BibitemOpen
  \bibfield  {author} {\bibinfo {author} {\bibfnamefont {K.~S.}\ \bibnamefont
  {Novoselov}}, \bibinfo {author} {\bibfnamefont {A.~K.}\ \bibnamefont {Geim}},
  \bibinfo {author} {\bibfnamefont {S.~V.}\ \bibnamefont {Morozov}}, \bibinfo
  {author} {\bibfnamefont {D.}~\bibnamefont {Jiang}}, \bibinfo {author}
  {\bibfnamefont {Y.}~\bibnamefont {Zhang}}, \bibinfo {author} {\bibfnamefont
  {S.~V.}\ \bibnamefont {Dubonos}}, \bibinfo {author} {\bibfnamefont {I.~V.}\
  \bibnamefont {Grigorieva}}, \ and\ \bibinfo {author} {\bibfnamefont {A.~A.}\
  \bibnamefont {Firsov}},\ }\href {\doibase 10.1126/science.1102896} {\bibfield
   {journal} {\bibinfo  {journal} {Science}\ }\textbf {\bibinfo {volume}
  {306}},\ \bibinfo {pages} {666} (\bibinfo {year} {2004})}\BibitemShut
  {NoStop}%
\bibitem [{\citenamefont {Tapaszt\'o}\ \emph {et~al.}(2008)\citenamefont
  {Tapaszt\'o}, \citenamefont {Dobrik}, \citenamefont {Lambin},\ and\
  \citenamefont {Bir\'o}}]{Tapaszto:litography}%
  \BibitemOpen
  \bibfield  {author} {\bibinfo {author} {\bibfnamefont {L.}~\bibnamefont
  {Tapaszt\'o}}, \bibinfo {author} {\bibfnamefont {G.}~\bibnamefont {Dobrik}},
  \bibinfo {author} {\bibfnamefont {P.}~\bibnamefont {Lambin}}, \ and\ \bibinfo
  {author} {\bibfnamefont {L.~P.}\ \bibnamefont {Bir\'o}},\ }\href
  {http://dx.doi.org/10.1038/nnano.2008.149} {\bibfield  {journal} {\bibinfo
  {journal} {Nat. Nano.}\ }\textbf {\bibinfo {volume} {3}},\ \bibinfo {pages}
  {397} (\bibinfo {year} {2008})}\BibitemShut {NoStop}%
\bibitem [{\citenamefont {Magda}\ \emph {et~al.}(2014)\citenamefont {Magda},
  \citenamefont {Jin}, \citenamefont {Hagym\'asi}, \citenamefont {Vancs\'o},
  \citenamefont {Osv\'ath}, \citenamefont {Nemes-Incze}, \citenamefont {Hwang},
  \citenamefont {Bir\'o},\ and\ \citenamefont {Tapaszt\'o}}]{Magda2014}%
  \BibitemOpen
  \bibfield  {author} {\bibinfo {author} {\bibfnamefont {G.~Z.}\ \bibnamefont
  {Magda}}, \bibinfo {author} {\bibfnamefont {X.}~\bibnamefont {Jin}}, \bibinfo
  {author} {\bibfnamefont {I.}~\bibnamefont {Hagym\'asi}}, \bibinfo {author}
  {\bibfnamefont {P.}~\bibnamefont {Vancs\'o}}, \bibinfo {author}
  {\bibfnamefont {Z.}~\bibnamefont {Osv\'ath}}, \bibinfo {author}
  {\bibfnamefont {P.}~\bibnamefont {Nemes-Incze}}, \bibinfo {author}
  {\bibfnamefont {C.}~\bibnamefont {Hwang}}, \bibinfo {author} {\bibfnamefont
  {L.~P.}\ \bibnamefont {Bir\'o}}, \ and\ \bibinfo {author} {\bibfnamefont
  {L.}~\bibnamefont {Tapaszt\'o}},\ }\href
  {http://dx.doi.org/10.1038/nature13831} {\bibfield  {journal} {\bibinfo
  {journal} {Nature}\ }\textbf {\bibinfo {volume} {514}},\ \bibinfo {pages}
  {608} (\bibinfo {year} {2014})}\BibitemShut {NoStop}%
\bibitem [{\citenamefont {Kimouche}\ \emph {et~al.}(2015)\citenamefont
  {Kimouche}, \citenamefont {Ervasti}, \citenamefont {Drost}, \citenamefont
  {Halonen}, \citenamefont {Harju}, \citenamefont {Joensuu}, \citenamefont
  {Sainio},\ and\ \citenamefont {Liljeroth}}]{Kimouche2015}%
  \BibitemOpen
  \bibfield  {author} {\bibinfo {author} {\bibfnamefont {A.}~\bibnamefont
  {Kimouche}}, \bibinfo {author} {\bibfnamefont {M.~M.}\ \bibnamefont
  {Ervasti}}, \bibinfo {author} {\bibfnamefont {R.}~\bibnamefont {Drost}},
  \bibinfo {author} {\bibfnamefont {S.}~\bibnamefont {Halonen}}, \bibinfo
  {author} {\bibfnamefont {A.}~\bibnamefont {Harju}}, \bibinfo {author}
  {\bibfnamefont {P.~M.}\ \bibnamefont {Joensuu}}, \bibinfo {author}
  {\bibfnamefont {J.}~\bibnamefont {Sainio}}, \ and\ \bibinfo {author}
  {\bibfnamefont {P.}~\bibnamefont {Liljeroth}},\ }\href
  {http://dx.doi.org/10.1038/ncomms10177} {\bibfield  {journal} {\bibinfo
  {journal} {Nat. Commun.}\ }\textbf {\bibinfo {volume} {6}},\ \bibinfo {pages}
  {10177} (\bibinfo {year} {2015})}\BibitemShut {NoStop}%
\bibitem [{\citenamefont {Ruffieux}\ \emph {et~al.}(2016)\citenamefont
  {Ruffieux}, \citenamefont {Wang}, \citenamefont {Yang}, \citenamefont
  {S{\'a}nchez-S{\'a}nchez}, \citenamefont {Liu}, \citenamefont {Dienel},
  \citenamefont {Talirz}, \citenamefont {Shinde}, \citenamefont {Pignedoli},
  \citenamefont {Passerone}, \citenamefont {Dumslaff}, \citenamefont {Feng},
  \citenamefont {M{\"u}llen},\ and\ \citenamefont {Fasel}}]{Ruffieux2016}%
  \BibitemOpen
  \bibfield  {author} {\bibinfo {author} {\bibfnamefont {P.}~\bibnamefont
  {Ruffieux}}, \bibinfo {author} {\bibfnamefont {S.}~\bibnamefont {Wang}},
  \bibinfo {author} {\bibfnamefont {B.}~\bibnamefont {Yang}}, \bibinfo {author}
  {\bibfnamefont {C.}~\bibnamefont {S{\'a}nchez-S{\'a}nchez}}, \bibinfo
  {author} {\bibfnamefont {J.}~\bibnamefont {Liu}}, \bibinfo {author}
  {\bibfnamefont {T.}~\bibnamefont {Dienel}}, \bibinfo {author} {\bibfnamefont
  {L.}~\bibnamefont {Talirz}}, \bibinfo {author} {\bibfnamefont
  {P.}~\bibnamefont {Shinde}}, \bibinfo {author} {\bibfnamefont {C.~A.}\
  \bibnamefont {Pignedoli}}, \bibinfo {author} {\bibfnamefont {D.}~\bibnamefont
  {Passerone}}, \bibinfo {author} {\bibfnamefont {T.}~\bibnamefont {Dumslaff}},
  \bibinfo {author} {\bibfnamefont {X.}~\bibnamefont {Feng}}, \bibinfo {author}
  {\bibfnamefont {K.}~\bibnamefont {M{\"u}llen}}, \ and\ \bibinfo {author}
  {\bibfnamefont {R.}~\bibnamefont {Fasel}},\ }\href
  {http://dx.doi.org/10.1038/nature17151} {\bibfield  {journal} {\bibinfo
  {journal} {Nature}\ }\textbf {\bibinfo {volume} {531}},\ \bibinfo {pages}
  {489} (\bibinfo {year} {2016})}\BibitemShut {NoStop}%
\bibitem [{\citenamefont {Fujita}\ \emph {et~al.}(1996)\citenamefont {Fujita},
  \citenamefont {Wakabayashi}, \citenamefont {Nakada},\ and\ \citenamefont
  {Kusakabe}}]{zz_edge_states}%
  \BibitemOpen
  \bibfield  {author} {\bibinfo {author} {\bibfnamefont {M.}~\bibnamefont
  {Fujita}}, \bibinfo {author} {\bibfnamefont {K.}~\bibnamefont {Wakabayashi}},
  \bibinfo {author} {\bibfnamefont {K.}~\bibnamefont {Nakada}}, \ and\ \bibinfo
  {author} {\bibfnamefont {K.}~\bibnamefont {Kusakabe}},\ }\href {\doibase
  10.1143/JPSJ.65.1920} {\bibfield  {journal} {\bibinfo  {journal} {J. Phys.
  Soc. Jpn.}\ }\textbf {\bibinfo {volume} {65}},\ \bibinfo {pages} {1920}
  (\bibinfo {year} {1996})}\BibitemShut {NoStop}%
\bibitem [{\citenamefont {Son}\ \emph {et~al.}(2006)\citenamefont {Son},
  \citenamefont {Cohen},\ and\ \citenamefont {Louie}}]{louie:prl2006}%
  \BibitemOpen
  \bibfield  {author} {\bibinfo {author} {\bibfnamefont {Y.-W.}\ \bibnamefont
  {Son}}, \bibinfo {author} {\bibfnamefont {M.~L.}\ \bibnamefont {Cohen}}, \
  and\ \bibinfo {author} {\bibfnamefont {S.~G.}\ \bibnamefont {Louie}},\ }\href
  {\doibase 10.1103/PhysRevLett.97.216803} {\bibfield  {journal} {\bibinfo
  {journal} {Phys. Rev. Lett.}\ }\textbf {\bibinfo {volume} {97}},\ \bibinfo
  {pages} {216803} (\bibinfo {year} {2006})}\BibitemShut {NoStop}%
\bibitem [{\citenamefont {Gonz\'alez}\ \emph {et~al.}(1999)\citenamefont
  {Gonz\'alez}, \citenamefont {Guinea},\ and\ \citenamefont
  {Vozmediano}}]{graphene:Fermi}%
  \BibitemOpen
  \bibfield  {author} {\bibinfo {author} {\bibfnamefont {J.}~\bibnamefont
  {Gonz\'alez}}, \bibinfo {author} {\bibfnamefont {F.}~\bibnamefont {Guinea}},
  \ and\ \bibinfo {author} {\bibfnamefont {M.~A.~H.}\ \bibnamefont
  {Vozmediano}},\ }\href {\doibase 10.1103/PhysRevB.59.R2474} {\bibfield
  {journal} {\bibinfo  {journal} {Phys. Rev. B}\ }\textbf {\bibinfo {volume}
  {59}},\ \bibinfo {pages} {R2474} (\bibinfo {year} {1999})}\BibitemShut
  {NoStop}%
\bibitem [{\citenamefont {Yang}\ \emph {et~al.}(2007)\citenamefont {Yang},
  \citenamefont {Park}, \citenamefont {Son}, \citenamefont {Cohen},\ and\
  \citenamefont {Louie}}]{Louie:GW}%
  \BibitemOpen
  \bibfield  {author} {\bibinfo {author} {\bibfnamefont {L.}~\bibnamefont
  {Yang}}, \bibinfo {author} {\bibfnamefont {C.-H.}\ \bibnamefont {Park}},
  \bibinfo {author} {\bibfnamefont {Y.-W.}\ \bibnamefont {Son}}, \bibinfo
  {author} {\bibfnamefont {M.~L.}\ \bibnamefont {Cohen}}, \ and\ \bibinfo
  {author} {\bibfnamefont {S.~G.}\ \bibnamefont {Louie}},\ }\href {\doibase
  10.1103/PhysRevLett.99.186801} {\bibfield  {journal} {\bibinfo  {journal}
  {Phys. Rev. Lett.}\ }\textbf {\bibinfo {volume} {99}},\ \bibinfo {pages}
  {186801} (\bibinfo {year} {2007})}\BibitemShut {NoStop}%
\bibitem [{\citenamefont {Wassmann}\ \emph {et~al.}(2008)\citenamefont
  {Wassmann}, \citenamefont {Seitsonen}, \citenamefont {Saitta}, \citenamefont
  {Lazzeri},\ and\ \citenamefont {Mauri}}]{PhysRevLett.101.096402}%
  \BibitemOpen
  \bibfield  {author} {\bibinfo {author} {\bibfnamefont {T.}~\bibnamefont
  {Wassmann}}, \bibinfo {author} {\bibfnamefont {A.~P.}\ \bibnamefont
  {Seitsonen}}, \bibinfo {author} {\bibfnamefont {A.~M.}\ \bibnamefont
  {Saitta}}, \bibinfo {author} {\bibfnamefont {M.}~\bibnamefont {Lazzeri}}, \
  and\ \bibinfo {author} {\bibfnamefont {F.}~\bibnamefont {Mauri}},\ }\href
  {\doibase 10.1103/PhysRevLett.101.096402} {\bibfield  {journal} {\bibinfo
  {journal} {Phys. Rev. Lett.}\ }\textbf {\bibinfo {volume} {101}},\ \bibinfo
  {pages} {096402} (\bibinfo {year} {2008})}\BibitemShut {NoStop}%
\bibitem [{\citenamefont {Yazyev}\ and\ \citenamefont
  {Katsnelson}(2008)}]{Katsnelson:DFT}%
  \BibitemOpen
  \bibfield  {author} {\bibinfo {author} {\bibfnamefont {O.~V.}\ \bibnamefont
  {Yazyev}}\ and\ \bibinfo {author} {\bibfnamefont {M.~I.}\ \bibnamefont
  {Katsnelson}},\ }\href {\doibase 10.1103/PhysRevLett.100.047209} {\bibfield
  {journal} {\bibinfo  {journal} {Phys. Rev. Lett.}\ }\textbf {\bibinfo
  {volume} {100}},\ \bibinfo {pages} {047209} (\bibinfo {year}
  {2008})}\BibitemShut {NoStop}%
\bibitem [{\citenamefont {Li}\ \emph {et~al.}(2013)\citenamefont {Li},
  \citenamefont {Zhou}, \citenamefont {Cabrera},\ and\ \citenamefont
  {Chen}}]{Chen:DFT}%
  \BibitemOpen
  \bibfield  {author} {\bibinfo {author} {\bibfnamefont {Y.}~\bibnamefont
  {Li}}, \bibinfo {author} {\bibfnamefont {Z.}~\bibnamefont {Zhou}}, \bibinfo
  {author} {\bibfnamefont {C.~R.}\ \bibnamefont {Cabrera}}, \ and\ \bibinfo
  {author} {\bibfnamefont {Z.}~\bibnamefont {Chen}},\ }\href@noop {} {\bibfield
   {journal} {\bibinfo  {journal} {Sci. Rep.}\ }\textbf {\bibinfo {volume}
  {3}},\ \bibinfo {pages} {2030} (\bibinfo {year} {2013})}\BibitemShut
  {NoStop}%
\bibitem [{\citenamefont {Fern\'andez-Rossier}(2008)}]{Rossier:HF}%
  \BibitemOpen
  \bibfield  {author} {\bibinfo {author} {\bibfnamefont {J.}~\bibnamefont
  {Fern\'andez-Rossier}},\ }\href {\doibase 10.1103/PhysRevB.77.075430}
  {\bibfield  {journal} {\bibinfo  {journal} {Phys. Rev. B}\ }\textbf {\bibinfo
  {volume} {77}},\ \bibinfo {pages} {075430} (\bibinfo {year}
  {2008})}\BibitemShut {NoStop}%
\bibitem [{\citenamefont {Yazyev}(2008)}]{Yazyev:prl2008}%
  \BibitemOpen
  \bibfield  {author} {\bibinfo {author} {\bibfnamefont {O.~V.}\ \bibnamefont
  {Yazyev}},\ }\href {\doibase 10.1103/PhysRevLett.101.037203} {\bibfield
  {journal} {\bibinfo  {journal} {Phys. Rev. Lett.}\ }\textbf {\bibinfo
  {volume} {101}},\ \bibinfo {pages} {037203} (\bibinfo {year}
  {2008})}\BibitemShut {NoStop}%
\bibitem [{\citenamefont {Jung}\ and\ \citenamefont
  {MacDonald}(2009)}]{MacDonald:prb2009}%
  \BibitemOpen
  \bibfield  {author} {\bibinfo {author} {\bibfnamefont {J.}~\bibnamefont
  {Jung}}\ and\ \bibinfo {author} {\bibfnamefont {A.~H.}\ \bibnamefont
  {MacDonald}},\ }\href {\doibase 10.1103/PhysRevB.79.235433} {\bibfield
  {journal} {\bibinfo  {journal} {Phys. Rev. B}\ }\textbf {\bibinfo {volume}
  {79}},\ \bibinfo {pages} {235433} (\bibinfo {year} {2009})}\BibitemShut
  {NoStop}%
\bibitem [{\citenamefont {Pi}\ \emph {et~al.}(2015)\citenamefont {Pi},
  \citenamefont {Dou}, \citenamefont {Tang},\ and\ \citenamefont
  {Kaun}}]{Pi:HF2015}%
  \BibitemOpen
  \bibfield  {author} {\bibinfo {author} {\bibfnamefont {S.-T.}\ \bibnamefont
  {Pi}}, \bibinfo {author} {\bibfnamefont {K.-P.}\ \bibnamefont {Dou}},
  \bibinfo {author} {\bibfnamefont {C.-S.}\ \bibnamefont {Tang}}, \ and\
  \bibinfo {author} {\bibfnamefont {C.-C.}\ \bibnamefont {Kaun}},\ }\href
  {\doibase http://dx.doi.org/10.1016/j.carbon.2015.06.069} {\bibfield
  {journal} {\bibinfo  {journal} {Carbon}\ }\textbf {\bibinfo {volume} {94}},\
  \bibinfo {pages} {196 } (\bibinfo {year} {2015})}\BibitemShut {NoStop}%
\bibitem [{\citenamefont {Jung}(2011)}]{Jung:HF}%
  \BibitemOpen
  \bibfield  {author} {\bibinfo {author} {\bibfnamefont {J.}~\bibnamefont
  {Jung}},\ }\href {\doibase 10.1103/PhysRevB.83.165415} {\bibfield  {journal}
  {\bibinfo  {journal} {Phys. Rev. B}\ }\textbf {\bibinfo {volume} {83}},\
  \bibinfo {pages} {165415} (\bibinfo {year} {2011})}\BibitemShut {NoStop}%
\bibitem [{\citenamefont {Jung}\ \emph {et~al.}(2009)\citenamefont {Jung},
  \citenamefont {Pereg-Barnea},\ and\ \citenamefont {MacDonald}}]{Jung:HFprl}%
  \BibitemOpen
  \bibfield  {author} {\bibinfo {author} {\bibfnamefont {J.}~\bibnamefont
  {Jung}}, \bibinfo {author} {\bibfnamefont {T.}~\bibnamefont {Pereg-Barnea}},
  \ and\ \bibinfo {author} {\bibfnamefont {A.~H.}\ \bibnamefont {MacDonald}},\
  }\href {\doibase 10.1103/PhysRevLett.102.227205} {\bibfield  {journal}
  {\bibinfo  {journal} {Phys. Rev. Lett.}\ }\textbf {\bibinfo {volume} {102}},\
  \bibinfo {pages} {227205} (\bibinfo {year} {2009})}\BibitemShut {NoStop}%
\bibitem [{\citenamefont {Kumazaki}\ and\ \citenamefont
  {Hirashima}(2008)}]{Hirashima:HF}%
  \BibitemOpen
  \bibfield  {author} {\bibinfo {author} {\bibfnamefont {H.}~\bibnamefont
  {Kumazaki}}\ and\ \bibinfo {author} {\bibfnamefont {D.~S.}\ \bibnamefont
  {Hirashima}},\ }\href {\doibase 10.1143/JPSJ.77.044705} {\bibfield  {journal}
  {\bibinfo  {journal} {J. Phys. Soc. Jpn.}\ }\textbf {\bibinfo {volume}
  {77}},\ \bibinfo {pages} {044705} (\bibinfo {year} {2008})}\BibitemShut
  {NoStop}%
\bibitem [{\citenamefont {Carvalho}\ \emph {et~al.}(2014)\citenamefont
  {Carvalho}, \citenamefont {Warnes},\ and\ \citenamefont
  {Lewenkopf}}]{Lewenkopf:HF}%
  \BibitemOpen
  \bibfield  {author} {\bibinfo {author} {\bibfnamefont {A.~R.}\ \bibnamefont
  {Carvalho}}, \bibinfo {author} {\bibfnamefont {J.~H.}\ \bibnamefont
  {Warnes}}, \ and\ \bibinfo {author} {\bibfnamefont {C.~H.}\ \bibnamefont
  {Lewenkopf}},\ }\href {\doibase 10.1103/PhysRevB.89.245444} {\bibfield
  {journal} {\bibinfo  {journal} {Phys. Rev. B}\ }\textbf {\bibinfo {volume}
  {89}},\ \bibinfo {pages} {245444} (\bibinfo {year} {2014})}\BibitemShut
  {NoStop}%
\bibitem [{\citenamefont {Feldner}\ \emph {et~al.}(2010)\citenamefont
  {Feldner}, \citenamefont {Meng}, \citenamefont {Honecker}, \citenamefont
  {Cabra}, \citenamefont {Wessel},\ and\ \citenamefont
  {Assaad}}]{Feldner:prb2010}%
  \BibitemOpen
  \bibfield  {author} {\bibinfo {author} {\bibfnamefont {H.}~\bibnamefont
  {Feldner}}, \bibinfo {author} {\bibfnamefont {Z.~Y.}\ \bibnamefont {Meng}},
  \bibinfo {author} {\bibfnamefont {A.}~\bibnamefont {Honecker}}, \bibinfo
  {author} {\bibfnamefont {D.}~\bibnamefont {Cabra}}, \bibinfo {author}
  {\bibfnamefont {S.}~\bibnamefont {Wessel}}, \ and\ \bibinfo {author}
  {\bibfnamefont {F.~F.}\ \bibnamefont {Assaad}},\ }\href {\doibase
  10.1103/PhysRevB.81.115416} {\bibfield  {journal} {\bibinfo  {journal} {Phys.
  Rev. B}\ }\textbf {\bibinfo {volume} {81}},\ \bibinfo {pages} {115416}
  (\bibinfo {year} {2010})}\BibitemShut {NoStop}%
\bibitem [{\citenamefont {Feldner}\ \emph {et~al.}(2011)\citenamefont
  {Feldner}, \citenamefont {Meng}, \citenamefont {Lang}, \citenamefont
  {Assaad}, \citenamefont {Wessel},\ and\ \citenamefont
  {Honecker}}]{Feldner:prl2011}%
  \BibitemOpen
  \bibfield  {author} {\bibinfo {author} {\bibfnamefont {H.}~\bibnamefont
  {Feldner}}, \bibinfo {author} {\bibfnamefont {Z.~Y.}\ \bibnamefont {Meng}},
  \bibinfo {author} {\bibfnamefont {T.~C.}\ \bibnamefont {Lang}}, \bibinfo
  {author} {\bibfnamefont {F.~F.}\ \bibnamefont {Assaad}}, \bibinfo {author}
  {\bibfnamefont {S.}~\bibnamefont {Wessel}}, \ and\ \bibinfo {author}
  {\bibfnamefont {A.}~\bibnamefont {Honecker}},\ }\href {\doibase
  10.1103/PhysRevLett.106.226401} {\bibfield  {journal} {\bibinfo  {journal}
  {Phys. Rev. Lett.}\ }\textbf {\bibinfo {volume} {106}},\ \bibinfo {pages}
  {226401} (\bibinfo {year} {2011})}\BibitemShut {NoStop}%
\bibitem [{\citenamefont {Golor}\ \emph {et~al.}(2014)\citenamefont {Golor},
  \citenamefont {Wessel},\ and\ \citenamefont {Schmidt}}]{Schmidt:QMC}%
  \BibitemOpen
  \bibfield  {author} {\bibinfo {author} {\bibfnamefont {M.}~\bibnamefont
  {Golor}}, \bibinfo {author} {\bibfnamefont {S.}~\bibnamefont {Wessel}}, \
  and\ \bibinfo {author} {\bibfnamefont {M.~J.}\ \bibnamefont {Schmidt}},\
  }\href {\doibase 10.1103/PhysRevLett.112.046601} {\bibfield  {journal}
  {\bibinfo  {journal} {Phys. Rev. Lett.}\ }\textbf {\bibinfo {volume} {112}},\
  \bibinfo {pages} {046601} (\bibinfo {year} {2014})}\BibitemShut {NoStop}%
\bibitem [{\citenamefont {Hikihara}\ \emph {et~al.}(2003)\citenamefont
  {Hikihara}, \citenamefont {Hu}, \citenamefont {Lin},\ and\ \citenamefont
  {Mou}}]{Hikihara:dmrg2003}%
  \BibitemOpen
  \bibfield  {author} {\bibinfo {author} {\bibfnamefont {T.}~\bibnamefont
  {Hikihara}}, \bibinfo {author} {\bibfnamefont {X.}~\bibnamefont {Hu}},
  \bibinfo {author} {\bibfnamefont {H.-H.}\ \bibnamefont {Lin}}, \ and\
  \bibinfo {author} {\bibfnamefont {C.-Y.}\ \bibnamefont {Mou}},\ }\href
  {\doibase 10.1103/PhysRevB.68.035432} {\bibfield  {journal} {\bibinfo
  {journal} {Phys. Rev. B}\ }\textbf {\bibinfo {volume} {68}},\ \bibinfo
  {pages} {035432} (\bibinfo {year} {2003})}\BibitemShut {NoStop}%
\bibitem [{\citenamefont {Mizukami}\ \emph {et~al.}(2013)\citenamefont
  {Mizukami}, \citenamefont {Kurashige},\ and\ \citenamefont
  {Yanai}}]{Mizukami:dmrg}%
  \BibitemOpen
  \bibfield  {author} {\bibinfo {author} {\bibfnamefont {W.}~\bibnamefont
  {Mizukami}}, \bibinfo {author} {\bibfnamefont {Y.}~\bibnamefont {Kurashige}},
  \ and\ \bibinfo {author} {\bibfnamefont {T.}~\bibnamefont {Yanai}},\ }\href
  {\doibase 10.1021/ct3008974} {\bibfield  {journal} {\bibinfo  {journal} {J.
  Chem. Theor. and Comp.}\ }\textbf {\bibinfo {volume} {9}},\ \bibinfo {pages}
  {401} (\bibinfo {year} {2013})},\ \bibinfo {note} {pMID:
  26589042}\BibitemShut {NoStop}%
\bibitem [{\citenamefont {Goli}\ \emph {et~al.}(2016)\citenamefont {Goli},
  \citenamefont {Prodhan}, \citenamefont {Mazumdar},\ and\ \citenamefont
  {Ramasesha}}]{Ramasesha:gnr}%
  \BibitemOpen
  \bibfield  {author} {\bibinfo {author} {\bibfnamefont {V.~M. L. D.~P.}\
  \bibnamefont {Goli}}, \bibinfo {author} {\bibfnamefont {S.}~\bibnamefont
  {Prodhan}}, \bibinfo {author} {\bibfnamefont {S.}~\bibnamefont {Mazumdar}}, \
  and\ \bibinfo {author} {\bibfnamefont {S.}~\bibnamefont {Ramasesha}},\ }\href
  {\doibase 10.1103/PhysRevB.94.035139} {\bibfield  {journal} {\bibinfo
  {journal} {Phys. Rev. B}\ }\textbf {\bibinfo {volume} {94}},\ \bibinfo
  {pages} {035139} (\bibinfo {year} {2016})}\BibitemShut {NoStop}%
\bibitem [{\citenamefont {Koop}\ and\ \citenamefont
  {Schmidt}(2015)}]{Schmidt:prb}%
  \BibitemOpen
  \bibfield  {author} {\bibinfo {author} {\bibfnamefont {C.}~\bibnamefont
  {Koop}}\ and\ \bibinfo {author} {\bibfnamefont {M.~J.}\ \bibnamefont
  {Schmidt}},\ }\href {\doibase 10.1103/PhysRevB.92.125416} {\bibfield
  {journal} {\bibinfo  {journal} {Phys. Rev. B}\ }\textbf {\bibinfo {volume}
  {92}},\ \bibinfo {pages} {125416} (\bibinfo {year} {2015})}\BibitemShut
  {NoStop}%
\bibitem [{\citenamefont {White}(1992)}]{White:DMRG1}%
  \BibitemOpen
  \bibfield  {author} {\bibinfo {author} {\bibfnamefont {S.~R.}\ \bibnamefont
  {White}},\ }\href@noop {} {\bibfield  {journal} {\bibinfo  {journal} {Phys.
  Rev. Lett.}\ }\textbf {\bibinfo {volume} {69}},\ \bibinfo {pages} {2863}
  (\bibinfo {year} {1992})}\BibitemShut {NoStop}%
\bibitem [{\citenamefont {White}(1993)}]{White:DMRG2}%
  \BibitemOpen
  \bibfield  {author} {\bibinfo {author} {\bibfnamefont {S.~R.}\ \bibnamefont
  {White}},\ }\href@noop {} {\bibfield  {journal} {\bibinfo  {journal} {Phys.
  Rev. B}\ }\textbf {\bibinfo {volume} {48}},\ \bibinfo {pages} {10345}
  (\bibinfo {year} {1993})}\BibitemShut {NoStop}%
\bibitem [{\citenamefont {Schollw\"ock}(2005)}]{schollwock2005}%
  \BibitemOpen
  \bibfield  {author} {\bibinfo {author} {\bibfnamefont {U.}~\bibnamefont
  {Schollw\"ock}},\ }\href@noop {} {\bibfield  {journal} {\bibinfo  {journal}
  {Rev. Mod. Phys.}\ }\textbf {\bibinfo {volume} {77}},\ \bibinfo {pages} {259}
  (\bibinfo {year} {2005})}\BibitemShut {NoStop}%
\bibitem [{\citenamefont {Noack}\ and\ \citenamefont
  {Manmana}(2005)}]{manmana2005}%
  \BibitemOpen
  \bibfield  {author} {\bibinfo {author} {\bibfnamefont {R.~M.}\ \bibnamefont
  {Noack}}\ and\ \bibinfo {author} {\bibfnamefont {S.}~\bibnamefont
  {Manmana}},\ }\href@noop {} {\bibfield  {journal} {\bibinfo  {journal} {AIP
  Conf. Proc.}\ }\textbf {\bibinfo {volume} {789}},\ \bibinfo {pages} {93}
  (\bibinfo {year} {2005})}\BibitemShut {NoStop}%
\bibitem [{\citenamefont {Hallberg}(2006)}]{hallberg2006}%
  \BibitemOpen
  \bibfield  {author} {\bibinfo {author} {\bibfnamefont {K.}~\bibnamefont
  {Hallberg}},\ }\href@noop {} {\bibfield  {journal} {\bibinfo  {journal} {Adv.
  Phys.}\ }\textbf {\bibinfo {volume} {55}},\ \bibinfo {pages} {477} (\bibinfo
  {year} {2006})}\BibitemShut {NoStop}%
\bibitem [{\citenamefont {Szalay}\ \emph {et~al.}(2015)\citenamefont {Szalay},
  \citenamefont {Pfeffer}, \citenamefont {Murg}, \citenamefont {Barcza},
  \citenamefont {Verstraete}, \citenamefont {Schneider},\ and\ \citenamefont
  {Legeza}}]{legeza:review}%
  \BibitemOpen
  \bibfield  {author} {\bibinfo {author} {\bibfnamefont {S.}~\bibnamefont
  {Szalay}}, \bibinfo {author} {\bibfnamefont {M.}~\bibnamefont {Pfeffer}},
  \bibinfo {author} {\bibfnamefont {V.}~\bibnamefont {Murg}}, \bibinfo {author}
  {\bibfnamefont {G.}~\bibnamefont {Barcza}}, \bibinfo {author} {\bibfnamefont
  {F.}~\bibnamefont {Verstraete}}, \bibinfo {author} {\bibfnamefont
  {R.}~\bibnamefont {Schneider}}, \ and\ \bibinfo {author} {\bibfnamefont
  {{\"O}.}~\bibnamefont {Legeza}},\ }\href {\doibase 10.1002/qua.24898}
  {\bibfield  {journal} {\bibinfo  {journal} {Int. J. Quant. Chem.}\ }\textbf
  {\bibinfo {volume} {115}},\ \bibinfo {pages} {1342} (\bibinfo {year}
  {2015})}\BibitemShut {NoStop}%
\bibitem [{\citenamefont {Legeza}\ \emph {et~al.}(2003)\citenamefont {Legeza},
  \citenamefont {R\"oder},\ and\ \citenamefont {Hess}}]{DBSS:cikk1}%
  \BibitemOpen
  \bibfield  {author} {\bibinfo {author} {\bibfnamefont {{\"O}.}~\bibnamefont
  {Legeza}}, \bibinfo {author} {\bibfnamefont {J.}~\bibnamefont {R\"oder}}, \
  and\ \bibinfo {author} {\bibfnamefont {B.~A.}\ \bibnamefont {Hess}},\
  }\href@noop {} {\bibfield  {journal} {\bibinfo  {journal} {Phys. Rev. B}\
  }\textbf {\bibinfo {volume} {67}},\ \bibinfo {pages} {125114} (\bibinfo
  {year} {2003})}\BibitemShut {NoStop}%
\bibitem [{\citenamefont {Legeza}\ and\ \citenamefont
  {S\'olyom}(2004{\natexlab{a}})}]{DBSS:cikk2}%
  \BibitemOpen
  \bibfield  {author} {\bibinfo {author} {\bibfnamefont {{\"O}.}~\bibnamefont
  {Legeza}}\ and\ \bibinfo {author} {\bibfnamefont {J.}~\bibnamefont
  {S\'olyom}},\ }\href@noop {} {\bibfield  {journal} {\bibinfo  {journal}
  {Phys. Rev. B}\ }\textbf {\bibinfo {volume} {70}},\ \bibinfo {pages} {205118}
  (\bibinfo {year} {2004}{\natexlab{a}})}\BibitemShut {NoStop}%
\bibitem [{\citenamefont {Lieb}(1989)}]{Lieb:prb}%
  \BibitemOpen
  \bibfield  {author} {\bibinfo {author} {\bibfnamefont {E.~H.}\ \bibnamefont
  {Lieb}},\ }\href {\doibase 10.1103/PhysRevLett.62.1201} {\bibfield  {journal}
  {\bibinfo  {journal} {Phys. Rev. Lett.}\ }\textbf {\bibinfo {volume} {62}},\
  \bibinfo {pages} {1201} (\bibinfo {year} {1989})}\BibitemShut {NoStop}%
\bibitem [{\citenamefont {Assaad}\ and\ \citenamefont
  {Herbut}(2013)}]{Mott1:prx}%
  \BibitemOpen
  \bibfield  {author} {\bibinfo {author} {\bibfnamefont {F.~F.}\ \bibnamefont
  {Assaad}}\ and\ \bibinfo {author} {\bibfnamefont {I.~F.}\ \bibnamefont
  {Herbut}},\ }\href {\doibase 10.1103/PhysRevX.3.031010} {\bibfield  {journal}
  {\bibinfo  {journal} {Phys. Rev. X}\ }\textbf {\bibinfo {volume} {3}},\
  \bibinfo {pages} {031010} (\bibinfo {year} {2013})}\BibitemShut {NoStop}%
\bibitem [{\citenamefont {Parisen~Toldin}\ \emph {et~al.}(2015)\citenamefont
  {Parisen~Toldin}, \citenamefont {Hohenadler}, \citenamefont {Assaad},\ and\
  \citenamefont {Herbut}}]{Mott2:prb}%
  \BibitemOpen
  \bibfield  {author} {\bibinfo {author} {\bibfnamefont {F.}~\bibnamefont
  {Parisen~Toldin}}, \bibinfo {author} {\bibfnamefont {M.}~\bibnamefont
  {Hohenadler}}, \bibinfo {author} {\bibfnamefont {F.~F.}\ \bibnamefont
  {Assaad}}, \ and\ \bibinfo {author} {\bibfnamefont {I.~F.}\ \bibnamefont
  {Herbut}},\ }\href {\doibase 10.1103/PhysRevB.91.165108} {\bibfield
  {journal} {\bibinfo  {journal} {Phys. Rev. B}\ }\textbf {\bibinfo {volume}
  {91}},\ \bibinfo {pages} {165108} (\bibinfo {year} {2015})}\BibitemShut
  {NoStop}%
\bibitem [{\citenamefont {Otsuka}\ \emph {et~al.}(2016)\citenamefont {Otsuka},
  \citenamefont {Yunoki},\ and\ \citenamefont {Sorella}}]{Mott3:prx}%
  \BibitemOpen
  \bibfield  {author} {\bibinfo {author} {\bibfnamefont {Y.}~\bibnamefont
  {Otsuka}}, \bibinfo {author} {\bibfnamefont {S.}~\bibnamefont {Yunoki}}, \
  and\ \bibinfo {author} {\bibfnamefont {S.}~\bibnamefont {Sorella}},\ }\href
  {\doibase 10.1103/PhysRevX.6.011029} {\bibfield  {journal} {\bibinfo
  {journal} {Phys. Rev. X}\ }\textbf {\bibinfo {volume} {6}},\ \bibinfo {pages}
  {011029} (\bibinfo {year} {2016})}\BibitemShut {NoStop}%
\bibitem [{\citenamefont {Wolf}\ \emph {et~al.}(2008)\citenamefont {Wolf},
  \citenamefont {Verstraete}, \citenamefont {Hastings},\ and\ \citenamefont
  {Cirac}}]{wolf2008}%
  \BibitemOpen
  \bibfield  {author} {\bibinfo {author} {\bibfnamefont {M.~M.}\ \bibnamefont
  {Wolf}}, \bibinfo {author} {\bibfnamefont {F.}~\bibnamefont {Verstraete}},
  \bibinfo {author} {\bibfnamefont {M.~B.}\ \bibnamefont {Hastings}}, \ and\
  \bibinfo {author} {\bibfnamefont {J.~I.}\ \bibnamefont {Cirac}},\ }\href@noop
  {} {\bibfield  {journal} {\bibinfo  {journal} {Phys. Rev. Lett.}\ }\textbf
  {\bibinfo {volume} {100}},\ \bibinfo {pages} {070502} (\bibinfo {year}
  {2008})}\BibitemShut {NoStop}%
\bibitem [{\citenamefont {Furukawa}\ \emph {et~al.}(2009)\citenamefont
  {Furukawa}, \citenamefont {Pasquier},\ and\ \citenamefont
  {Shiraishi}}]{furukawa2009}%
  \BibitemOpen
  \bibfield  {author} {\bibinfo {author} {\bibfnamefont {S.}~\bibnamefont
  {Furukawa}}, \bibinfo {author} {\bibfnamefont {V.}~\bibnamefont {Pasquier}},
  \ and\ \bibinfo {author} {\bibfnamefont {J.}~\bibnamefont {Shiraishi}},\
  }\href@noop {} {\bibfield  {journal} {\bibinfo  {journal} {Phys. Rev. Lett.}\
  }\textbf {\bibinfo {volume} {102}},\ \bibinfo {pages} {170602} (\bibinfo
  {year} {2009})}\BibitemShut {NoStop}%
\bibitem [{\citenamefont {Barcza}\ \emph {et~al.}(2015)\citenamefont {Barcza},
  \citenamefont {Noack}, \citenamefont {S\'olyom},\ and\ \citenamefont
  {Legeza}}]{legeza:entanglement}%
  \BibitemOpen
  \bibfield  {author} {\bibinfo {author} {\bibfnamefont {G.}~\bibnamefont
  {Barcza}}, \bibinfo {author} {\bibfnamefont {R.~M.}\ \bibnamefont {Noack}},
  \bibinfo {author} {\bibfnamefont {J.}~\bibnamefont {S\'olyom}}, \ and\
  \bibinfo {author} {\bibfnamefont {{\"O}.}~\bibnamefont {Legeza}},\ }\href
  {\doibase 10.1103/PhysRevB.92.125140} {\bibfield  {journal} {\bibinfo
  {journal} {Phys. Rev. B}\ }\textbf {\bibinfo {volume} {92}},\ \bibinfo
  {pages} {125140} (\bibinfo {year} {2015})}\BibitemShut {NoStop}%
\bibitem [{\citenamefont {Legeza}\ and\ \citenamefont
  {S\'olyom}(2003)}]{legeza2003b}%
  \BibitemOpen
  \bibfield  {author} {\bibinfo {author} {\bibfnamefont {{\"O}.}~\bibnamefont
  {Legeza}}\ and\ \bibinfo {author} {\bibfnamefont {J.}~\bibnamefont
  {S\'olyom}},\ }\href@noop {} {\bibfield  {journal} {\bibinfo  {journal}
  {Phys. Rev. B}\ }\textbf {\bibinfo {volume} {68}},\ \bibinfo {pages} {195116}
  (\bibinfo {year} {2003})}\BibitemShut {NoStop}%
\bibitem [{\citenamefont {Vidal}\ \emph {et~al.}(2003)\citenamefont {Vidal},
  \citenamefont {Latorre}, \citenamefont {Rico},\ and\ \citenamefont
  {Kitaev}}]{vidallatorre03}%
  \BibitemOpen
  \bibfield  {author} {\bibinfo {author} {\bibfnamefont {G.}~\bibnamefont
  {Vidal}}, \bibinfo {author} {\bibfnamefont {J.~I.}\ \bibnamefont {Latorre}},
  \bibinfo {author} {\bibfnamefont {E.}~\bibnamefont {Rico}}, \ and\ \bibinfo
  {author} {\bibfnamefont {A.}~\bibnamefont {Kitaev}},\ }\href@noop {}
  {\bibfield  {journal} {\bibinfo  {journal} {Phys. Rev. Lett.}\ }\textbf
  {\bibinfo {volume} {90}},\ \bibinfo {pages} {227902} (\bibinfo {year}
  {2003})}\BibitemShut {NoStop}%
\bibitem [{\citenamefont {Calabrese}\ and\ \citenamefont
  {Cardy}(2004)}]{calabrese04}%
  \BibitemOpen
  \bibfield  {author} {\bibinfo {author} {\bibfnamefont {P.}~\bibnamefont
  {Calabrese}}\ and\ \bibinfo {author} {\bibfnamefont {J.}~\bibnamefont
  {Cardy}},\ }\href@noop {} {\bibfield  {journal} {\bibinfo  {journal} {J.
  Stat. Mech.}\ }\textbf {\bibinfo {volume} {2004}},\ \bibinfo {pages} {P06002}
  (\bibinfo {year} {2004})}\BibitemShut {NoStop}%
\bibitem [{\citenamefont {Legeza}\ and\ \citenamefont
  {S\'olyom}(2006)}]{legeza2006}%
  \BibitemOpen
  \bibfield  {author} {\bibinfo {author} {\bibfnamefont {{\"O}.}~\bibnamefont
  {Legeza}}\ and\ \bibinfo {author} {\bibfnamefont {J.}~\bibnamefont
  {S\'olyom}},\ }\href@noop {} {\bibfield  {journal} {\bibinfo  {journal}
  {Phys. Rev. Lett.}\ }\textbf {\bibinfo {volume} {96}},\ \bibinfo {pages}
  {116401} (\bibinfo {year} {2006})}\BibitemShut {NoStop}%
\bibitem [{\citenamefont {Rissler}\ \emph {et~al.}(2006)\citenamefont
  {Rissler}, \citenamefont {Noack},\ and\ \citenamefont {White}}]{rissler2006}%
  \BibitemOpen
  \bibfield  {author} {\bibinfo {author} {\bibfnamefont {J.}~\bibnamefont
  {Rissler}}, \bibinfo {author} {\bibfnamefont {R.~M.}\ \bibnamefont {Noack}},
  \ and\ \bibinfo {author} {\bibfnamefont {S.~R.}\ \bibnamefont {White}},\
  }\href@noop {} {\bibfield  {journal} {\bibinfo  {journal} {Chem. Phys.}\
  }\textbf {\bibinfo {volume} {323}},\ \bibinfo {pages} {519 } (\bibinfo {year}
  {2006})}\BibitemShut {NoStop}%
\bibitem [{\citenamefont {Amico}\ \emph {et~al.}(2008)\citenamefont {Amico},
  \citenamefont {Fazio}, \citenamefont {Osterloh},\ and\ \citenamefont
  {Vedral}}]{luigi2008}%
  \BibitemOpen
  \bibfield  {author} {\bibinfo {author} {\bibfnamefont {L.}~\bibnamefont
  {Amico}}, \bibinfo {author} {\bibfnamefont {R.}~\bibnamefont {Fazio}},
  \bibinfo {author} {\bibfnamefont {A.}~\bibnamefont {Osterloh}}, \ and\
  \bibinfo {author} {\bibfnamefont {V.}~\bibnamefont {Vedral}},\ }\href@noop {}
  {\bibfield  {journal} {\bibinfo  {journal} {Rev. Mod. Phys.}\ }\textbf
  {\bibinfo {volume} {80}},\ \bibinfo {pages} {517} (\bibinfo {year}
  {2008})}\BibitemShut {NoStop}%
\bibitem [{\citenamefont {Legeza}\ and\ \citenamefont
  {S\'olyom}(2004{\natexlab{b}})}]{legeza2004}%
  \BibitemOpen
  \bibfield  {author} {\bibinfo {author} {\bibfnamefont {{\"O}.}~\bibnamefont
  {Legeza}}\ and\ \bibinfo {author} {\bibfnamefont {J.}~\bibnamefont
  {S\'olyom}},\ }\href {\doibase 10.1103/PhysRevB.70.205118} {\bibfield
  {journal} {\bibinfo  {journal} {Phys. Rev. B}\ }\textbf {\bibinfo {volume}
  {70}},\ \bibinfo {pages} {205118} (\bibinfo {year}
  {2004}{\natexlab{b}})}\BibitemShut {NoStop}%
\bibitem [{\citenamefont {Wehling}\ \emph {et~al.}(2011)\citenamefont
  {Wehling}, \citenamefont {\ifmmode \mbox{\c{S}}\else \c{S}\fi{}a\ifmmode
  \mbox{\c{s}}\else \c{s}\fi{}\ifmmode \imath \else \i
  \fi{}o\ifmmode~\breve{g}\else \u{g}\fi{}lu}, \citenamefont {Friedrich},
  \citenamefont {Lichtenstein}, \citenamefont {Katsnelson},\ and\ \citenamefont
  {Bl\"ugel}}]{graphene:abinitio}%
  \BibitemOpen
  \bibfield  {author} {\bibinfo {author} {\bibfnamefont {T.~O.}\ \bibnamefont
  {Wehling}}, \bibinfo {author} {\bibfnamefont {E.}~\bibnamefont {\ifmmode
  \mbox{\c{S}}\else \c{S}\fi{}a\ifmmode \mbox{\c{s}}\else \c{s}\fi{}\ifmmode
  \imath \else \i \fi{}o\ifmmode~\breve{g}\else \u{g}\fi{}lu}}, \bibinfo
  {author} {\bibfnamefont {C.}~\bibnamefont {Friedrich}}, \bibinfo {author}
  {\bibfnamefont {A.~I.}\ \bibnamefont {Lichtenstein}}, \bibinfo {author}
  {\bibfnamefont {M.~I.}\ \bibnamefont {Katsnelson}}, \ and\ \bibinfo {author}
  {\bibfnamefont {S.}~\bibnamefont {Bl\"ugel}},\ }\href {\doibase
  10.1103/PhysRevLett.106.236805} {\bibfield  {journal} {\bibinfo  {journal}
  {Phys. Rev. Lett.}\ }\textbf {\bibinfo {volume} {106}},\ \bibinfo {pages}
  {236805} (\bibinfo {year} {2011})}\BibitemShut {NoStop}%
\bibitem [{\citenamefont {Kharche}\ and\ \citenamefont
  {Meunier}(2016)}]{dft:substrate}%
  \BibitemOpen
  \bibfield  {author} {\bibinfo {author} {\bibfnamefont {N.}~\bibnamefont
  {Kharche}}\ and\ \bibinfo {author} {\bibfnamefont {V.}~\bibnamefont
  {Meunier}},\ }\href {\doibase 10.1021/acs.jpclett.6b00422} {\bibfield
  {journal} {\bibinfo  {journal} {J. Phys. Chem. Lett.}\ }\textbf {\bibinfo
  {volume} {7}},\ \bibinfo {pages} {1526} (\bibinfo {year} {2016})}\BibitemShut
  {NoStop}%
\bibitem [{\citenamefont {Sorella}\ and\ \citenamefont
  {Tosatti}(1992)}]{Tosatti:epl}%
  \BibitemOpen
  \bibfield  {author} {\bibinfo {author} {\bibfnamefont {S.}~\bibnamefont
  {Sorella}}\ and\ \bibinfo {author} {\bibfnamefont {E.}~\bibnamefont
  {Tosatti}},\ }\href {http://stacks.iop.org/0295-5075/19/i=8/a=007} {\bibfield
   {journal} {\bibinfo  {journal} {EPL (Europhysics Letters)}\ }\textbf
  {\bibinfo {volume} {19}},\ \bibinfo {pages} {699} (\bibinfo {year}
  {1992})}\BibitemShut {NoStop}%
\end{thebibliography}%
%\printbibliography

\end{document}